\documentclass[12pt]{article}
\usepackage{fullpage, natbib, authblk, hyperref}
\usepackage{amsfonts, amsmath, amssymb, amsthm, amsbsy, bm, bbm}
\usepackage{rotating, caption, subcaption, wrapfig, multirow, graphicx}
\usepackage[usenames,dvipsnames]{xcolor}
\usepackage{floatrow}
\title{Bayesian estimation of in-game home team win probability for National Basketball Association games}
\author[1]{Jason T.~Maddox}
\author[2]{Ryan Sides}
\author[3]{Jane L.~Harvill}
\affil[1]{Department of Sport Management, Syracuse University}
\affil[2]{Department of Mathematics and Computer Science, Texas Woman's University} 
\affil[3]{Department of Statistical Science, Statistical Science, Baylor University}
\date{\today}
\begin{document}
\maketitle
\newpage
\begin{abstract}
\citet{maddoxetal_2022} establish a new win probability estimation for college basketball and compared the results with previous methods of \citet{stern_1994}, \citet{deshpande_jensen_2016}, and \citet{benz2019}.  This paper proposes modifications to the approach of~\citeauthor{maddoxetal_2022} for the NBA game and investigate the performance of the model.  Enhancements to the model are developed, and the resulting adjusted model is compared with existing methods and to the ESPN counterpart.  To illustrate utility, all methods are applied to the November 23, 2019 game between the Chicago Bulls and Charlotte Hornets.
\end{abstract}


\section{Introduction}

Sports analytics has become a well-established area of research.  Work spans more than 30 years and a large range of difficulty.  Since the early 2000s, research in statistical methods for sports analytics has risen dramatically.  The review articles of~\citet{Kubatko_etal_2007}, \citet{fernandez_2019}, and~\citet{terner_franks_2020} provide a fairly comprehensive review for sports analytics for a wide variety of sports, including basketball.  One problem of interest is predicting the probability that the home team wins during the course of the game, or predicting ``in-game win probability.'' 

Speaking broadly, models for predicting the outcome of a sporting event can be classified into two systems: (1) pregame prediction or (2) in-game, or in-play, prediction.  pregame prediction involves determining the outcome of a game before play begins.  Once play begins, the process of predicting the outcome ends.  In contrast, in-game prediction attempts to use the progress during a game to determine win probabilities that vary as a function of in-game variables, for example, elapsed game time or score difference.  The focus of this paper is in-game prediction.  

For a variety of sports, there are many different methods for accurately estimating in-game win probability found in the literature.  For basketball, one of the first attempts to estimate in-game basketball analytics is~\citet{westfall_1990} who developed a graphical summary of the scoring activity for a basketball game that is a real-time plot of the score difference versus the elapsed time.  The features of the graph provided easy access to largest leads, lead changes, come-from-behind activity, and other interesting game features.  As computing technology and algorithms have become more sophisticated, models for forecasting in-game win probability have become more complex.  Some are built on expert predictions, some on betting paradigms, and others on within-game metrics.  \citet{shirley_2007} modeled a basketball game using a Markov model with three states and used that model for estimating in-game win probability.  \citet{strumbelj_vracar_2012} improved upon that by taking into consideration the strengths of the two teams and estimated transition probabilities using performance statistics.  \citet{vracar-etal-2016} extended the state description to capture other facets beyond in-game states so that the transition probabilities become conditional on a broader game context. \citet{bashuk_2012} proposed using cumulative win probabilities over the duration of a game to measure team performance.  Using five years of game play, he generated a win probability index for NCAA basketball.  Using the index, he created an open system to measure the impact, in terms of win probability added, of each play.  More recently, \citet{benz2019} developed a logistic regression approach where the coefficients of the covariates are allowed to change a function of time, i.e., the effects of the coefficients are dynamic in nature.  \citet{chen-fan-2018} developed a method for modeling point differences using a functional data approach.  \citet{shi_song_2019} develop a discrete-time, finite-state Markov model for the progress of basketball scores, and use it to conditionally predict the probability the home team wins or loses by a certain amount.  \citet{song_shi_2020} present an in-play prediction model based on the gamma process.  They apply a Bayesian dynamic forecasting procedure that can be used to predict the final score and total points.  \citet{song-gao-shi-2020} modify the gamma process by employing betting lines, letting the expectation of the final points total equal the pregame betting line.  \citet{maddoxetal_2022} develop three Bayesian approaches with dynamic priors.  The adjusted model with a dynamic prior is, overall, the better of their three proposed methods.  In this paper, we adopt the approach of~\citeauthor{maddoxetal_2022} for predicting in-game win probabilities for games in the National Basketball Association (NBA). Additional considerations are made to their methodology upon the extension to the NBA, such as an refinement of the prior, improvement of the binning method and a more sophisticated adjustment from pregame information into the model for estimating in-game win probabilities. In addition, the win probabilities that ESPN publishes on their website have been collected to compare to the methodology proposed by \citeauthor{maddoxetal_2022}.

The remainder of the paper is organized as follows.  Section~\ref{sec:datacollection} provides a brief description of the data and the data collection process.    Section~\ref{sec:model} contains an more thorough explanation of the dynamic Bayesian estimator in~\cite{maddoxetal_2022} and describes the adjustments necessary to implement their approach to games in the NBA.  Section~\ref{sec:adjbayes} presents the adjusted dynamic Bayesian method for estimating in-game home team win probabilities, along with a comparison of performances of the dynamic Bayesian and adjusted dynamic Bayesian estimators and the ESPN counterpart.  To illustrate utility, Section~\ref{sec:application}  applies the Bayesian approaches to a specific NBA game, and compares that to the ESPN counterpart.  The paper concludes with a summary in~Section~\ref{sec:conclusion}.

\section{Data Collection}
\label{sec:datacollection}
The primary goal of the models that developed within is to find a practical approach for effectively predicting regular season in-game home team win probability for a single NBA game or a collection of NBA games.  The data collected for investigating the proposed models performances were taken from ESPN.  Specifically, play-by-play data from ESPN was scraped using \emph{R} \citep{R} and the package {\tt rvest} \citep{rvest}. The data was collected starting with the beginning of the 2012-13 NBA season through the 2019-20 season, when play was halted due to the COVID-19 pandemic. Due to the unprecedented and unpredictable nature of playing the end of the 2019-20 season inside of a ``bubble'' with all games on a neutral court, along with attendance limitations and inconsistent schedules for the 2020-21 season, the end of the 2019-20 season and the entire 2020-21 season are omitted. The postseason is also not included, since it is conceivable that postseason games have different behavior than regular season games. 

Along with collecting the play-by-play data from ESPN, ESPN has their own win probability model. Their model for estimating in-game win probability is not accessible, and may be considered a ``black box'' model.  The predicted win probabilities for the 2018-19 and 2019-20 seasons are available on ESPN's API that can be accessed through the ESPN Developer Center.\footnote{The description of the ESPN Developer Center may be found at \url{www.espn.com/apis/devcenter/overview.html}.} 
There is a general form of the URL for NBA game back-end data providing access to the win probabilities.  For each game, the game id changes in the URL. The game ids are scraped for all games in each season, then input into this URL to scrape the win probabilities throughout the game.

\section{Dynamic Bayesian Estimator}
\label{sec:model}
For a specific game, consider the random process that is the home team's lead at time $t = 0, 1, \ldots, 2879$, where $t$ is the game time elapsed in seconds.  At a specific time $t$ and for a specific home team lead $\ell$, let $p_{t, \ell}$ denote the in-game probability that the home team will win the game at the end of regulation. When considering multiple games $i = 1, 2, \ldots, M$, let $Y_i = 1$ if the home team wins game $i$ and 0 otherwise.  Consider $p_{t, \ell}$ as a continuous function of $t$ and $\ell$. \citet{maddoxetal_2022} establish an estimator of $p_{t, \ell}$ using a Bayesian approach. For each cell, they combine the data with a beta$(\alpha_{t, \ell}, \beta_{t, \ell})$ prior where $\alpha_{t, \ell}$ and $\beta_{t, \ell}$ are chosen based on $t$ and $\ell$.    Specifically, the game is partitioned into cells based on time (in seconds) and point differential.  Some $(t, \ell)$ cells may have small values of $N_{t, \ell} =$ number of games in that cell, which have the potential to result in estimators of $p_{t,\ell}$ with very large standard errors.  To address this issue, windows can be defined and centered on $(t,\ell)$ in such a way that the in-game win probability remains relatively constant across the window.  In basketball, since no offensive possession can result in more than four points the window with respect to $\ell$ can be reasonably defined as $[\ell - 2, \ell + 2]$.  Moreover, since most offensive possessions last at least six seconds the width of the time window is taken to be six.  The same notation will be adopted for any $[t - 3, t + 3] \times [\ell - 2, \ell + 2]$ window; that is, $N_{t,\ell}$ is the number of games in the window in which the home team has led by any value in $[\ell - 2, \ell + 2]$ points after any time in $[t - 3, t + 3]$ seconds and $n_{t,\ell} = \sum_{i=1}^{N_{t,\ell}} Y_i$, distributed as a binomial($N_{t,\ell}, p_{t,\ell})$ random variable.  Based on the binomial distribution, a simple estimator for in-game home team win probability for for each $(t, \ell)$ combination is the maximum likelihood estimator
\begin{equation}
    \label{eq:MLE}
    \bar p_{t, \ell} = \frac{n_{t, \ell}}{N_{t, \ell}}.
\end{equation}
In Section~\ref{sec:bins}, limitations of this binning method are explained and addressed.  

Then~\citeauthor{maddoxetal_2022} apply a Bayesian methods approach to estimate $p_{t,\ell}$.  Since the beta family of distributions is a conjugate prior for the binomial distribution, the beta-binomial connection is used to estimate $p_{t, \ell}$.  Let $\alpha_{t,\ell} > 0$ and $\beta_{t,\ell} > 0$ be the shape parameters of the beta prior on $p_{t,\ell}$.  Then for each window within the $(t, \ell)$ plane, the Bayes estimator of $p_{t,\ell}$ is the mean of the posterior beta distribution, specifically
\begin{equation}
    \label{eq:bayes}
    \hat p_{t,\ell} = \frac{n_{t,\ell} + \alpha_{t,\ell}}{N_{t,\ell} + \alpha_{t,\ell} + \beta_{t,\ell}}.
\end{equation}
In~\citeauthor{maddoxetal_2022}, the choice of $\alpha_{t, \ell}$ and $\beta_{t,\ell}$ are dependent both on the time remaining in the game and the score differential, leading to a ``dynamic prior.''  For the NBA dynamic Bayesian estimator, the choice of parameters for the beta prior is described in Section~\ref{sec:dynpriorNBA}.

\subsection{Dynamic Prior for NBA Games}
\label{sec:dynpriorNBA}
For the NBA, a more precise prior structure for the structure of the prior distribution is proposed compared to~\citet{maddoxetal_2022}.  A sample of 14 NBA experts, including NBA front office associates were polled.   For each combination of elapsed time and home team lead in Table~\ref{tab:dynamicprior}, the experts provided their estimation of the probability of a team winning, regardless of home team.  The sample mean $\tilde p_{t, \ell}$ and sample variance $s^2_{t,\ell}$ of the probabilities were computed.  The two scale parameters were estimated via a method-of-moments type approach.  The system of equations
\begin{align*}
    \tilde p_{t, \ell} & = \frac{\alpha_{t, \ell}}{\alpha_{t, \ell} + \beta_{t, \ell}}, \\
    s^2_{t, \ell} & = \frac{\alpha_{t, \ell}\beta_{t, \ell}}{\left(\alpha_{t, \ell} + \beta_{t, \ell}\right)^2\left(\alpha_{t, \ell} + \beta_{t, \ell} + 1\right)},
\end{align*}
is solved for $\alpha_{t, \ell}$ and $\beta_{t, \ell}$, yielding
\begin{align*}
    \alpha_{t, \ell} & = -\frac{\tilde p_{t, \ell}\left(\tilde p^2_{t,\ell} - \tilde p_{t, \ell} + s^2_{t, \ell}\right)}{s^2_{t, \ell}} \\
    \beta_{t, \ell} & = \frac{\left(\tilde p_{t, \ell} - 1\right)\left(\tilde p^2_{t, \ell} - \tilde p_{t, \ell} + s^2_{t, \ell}\right)}{s^2_{t, \ell}},
\end{align*}
as long as $(\tilde p_{t, \ell} - 1)\tilde p_{t, \ell}(\tilde p_{t, \ell}^2 - \tilde p_{t, \ell} + s^2_{t, \ell}) \ne 0$.

\begin{figure}[!ht]
\caption{Densities of beta(19, 7) prior (in blue) and beta (7, 19) prior (in red).}
\label{fig:betapriors}
\centering
\includegraphics[width=0.6\textwidth]{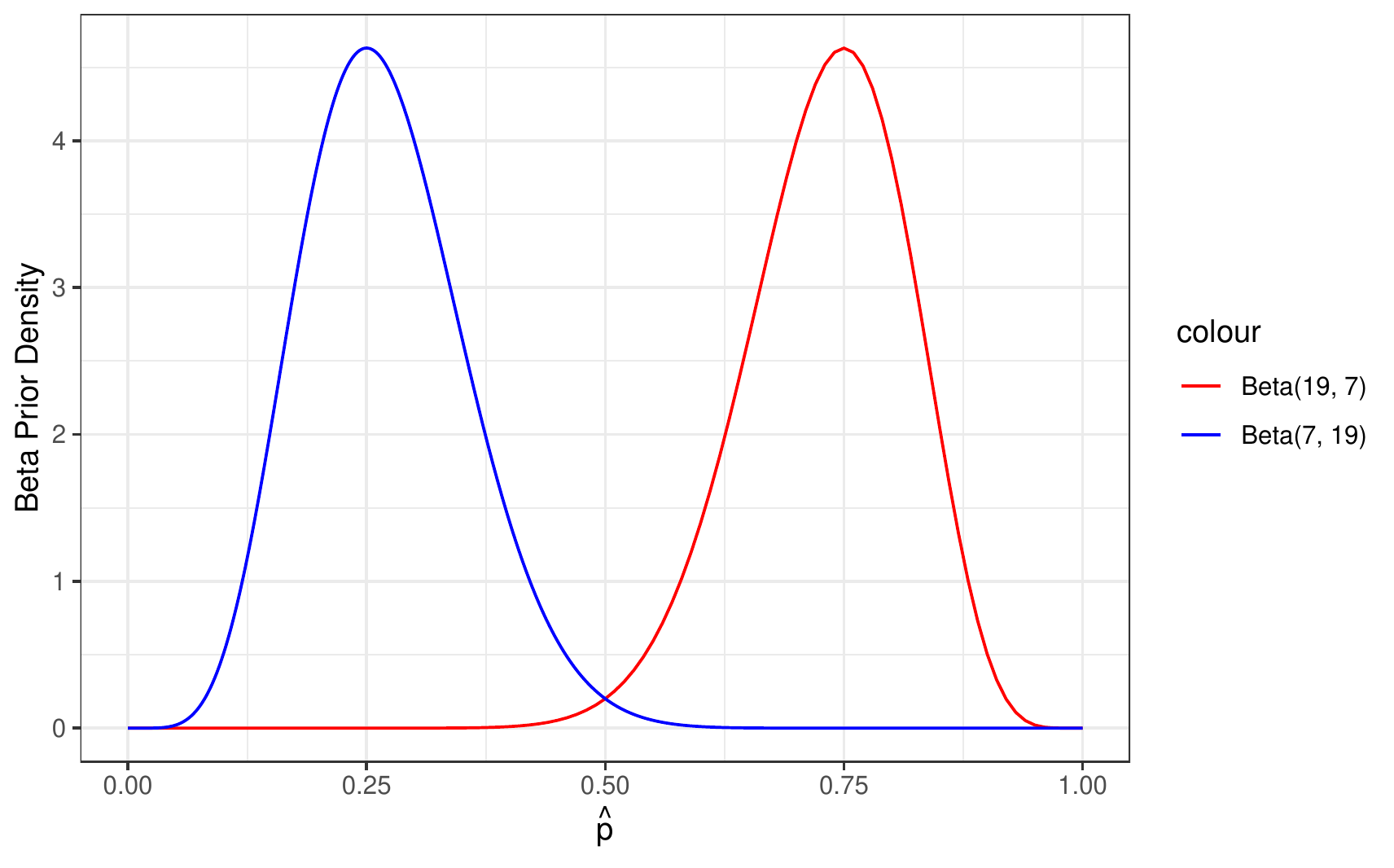}
\end{figure}

In Table~\ref{tab:dynamicprior}, the score differential is presented when the home team has the lead.  The greater the lead, the more likely the home team will win, and this is modeled with a left-skewed prior.  On the other hand, if the visiting team has the lead, then the roles of $\alpha_{t, \ell}$ and $\beta_{t, \ell}$ are reversed, and the prior becomes right-skewed.  The prior densities in Figure~\ref{fig:betapriors} illustrate this principle.  The curves are beta density priors plotted by home-team win probability.  For example, the fifth row in Table~\ref{tab:dynamicprior} for $t = 360, 361, \ldots, 719$ seconds and the home team has a lead of $\ell = 10, 11, \ldots, 19$ points has a beta(19, 7) prior seen in the figure as the blue density curve.  On the other hand, if the away team as the same lead, then the prior is a beta(7, 19), represented by the red curve.  At any time, if the game is tied ($\ell = 0$), or is sufficiently close in score for the amount of time remaining, the prior distribution is flat, or ``uninformative,'' meaning it gives equal weight to both teams winning the game.

\begin{figure}[!ht]
  \caption{Plots illustrating estimated in-game home team win probabilities based on maximum likelihood and Bayes with dynamic prior.}
  \begin{subfigure}[t]{0.475\textwidth}
\centering    
    \caption{Maximum likelihood estimates.} 
    \includegraphics[width=\textwidth]{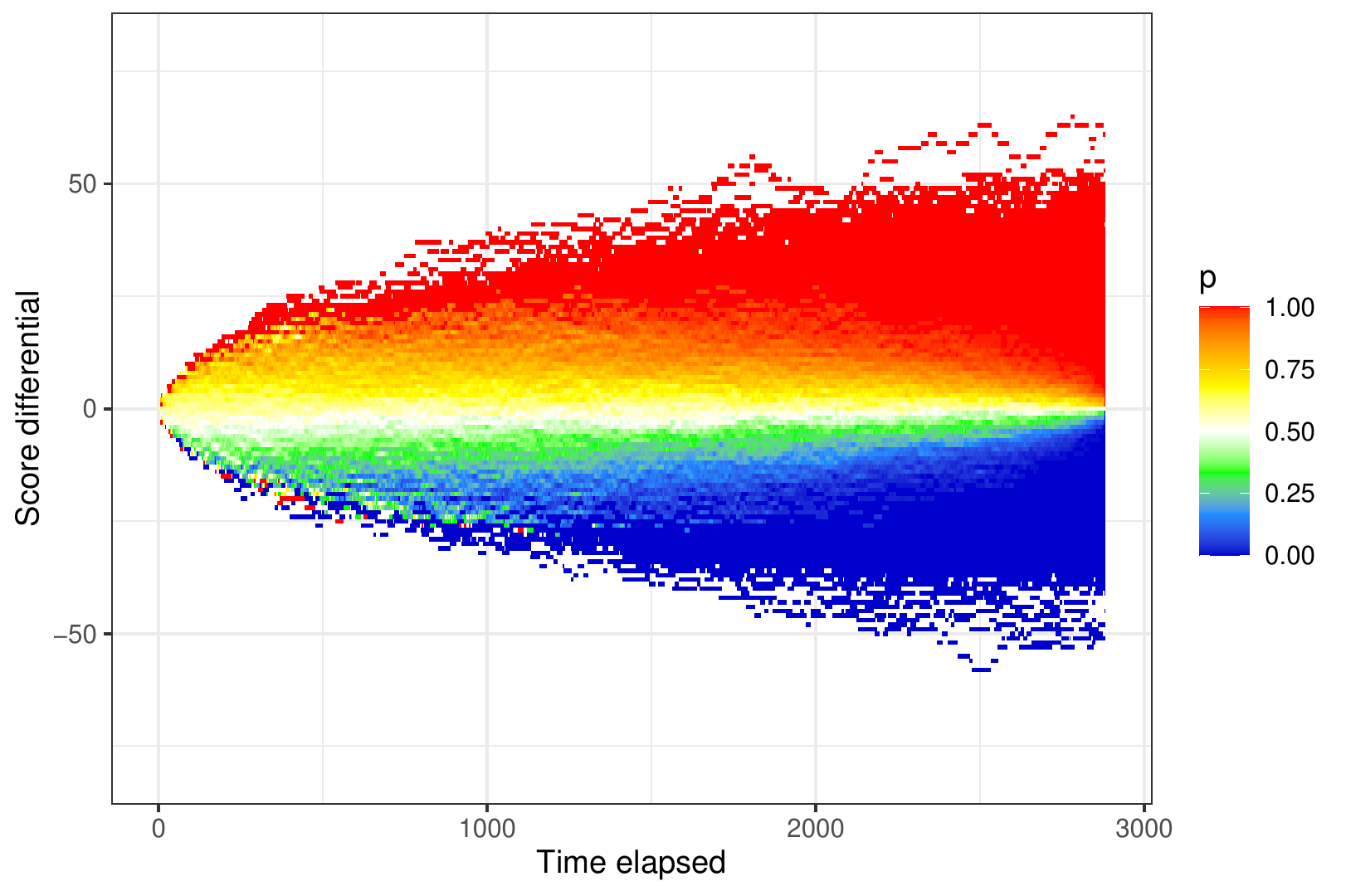}
    \label{fig:MLE}
  \end{subfigure}
  \hfill
  \begin{subfigure}[t]{0.475\textwidth}
   \centering
    \caption{Dynamic Bayesian estimates.}
    \includegraphics[width=\textwidth]{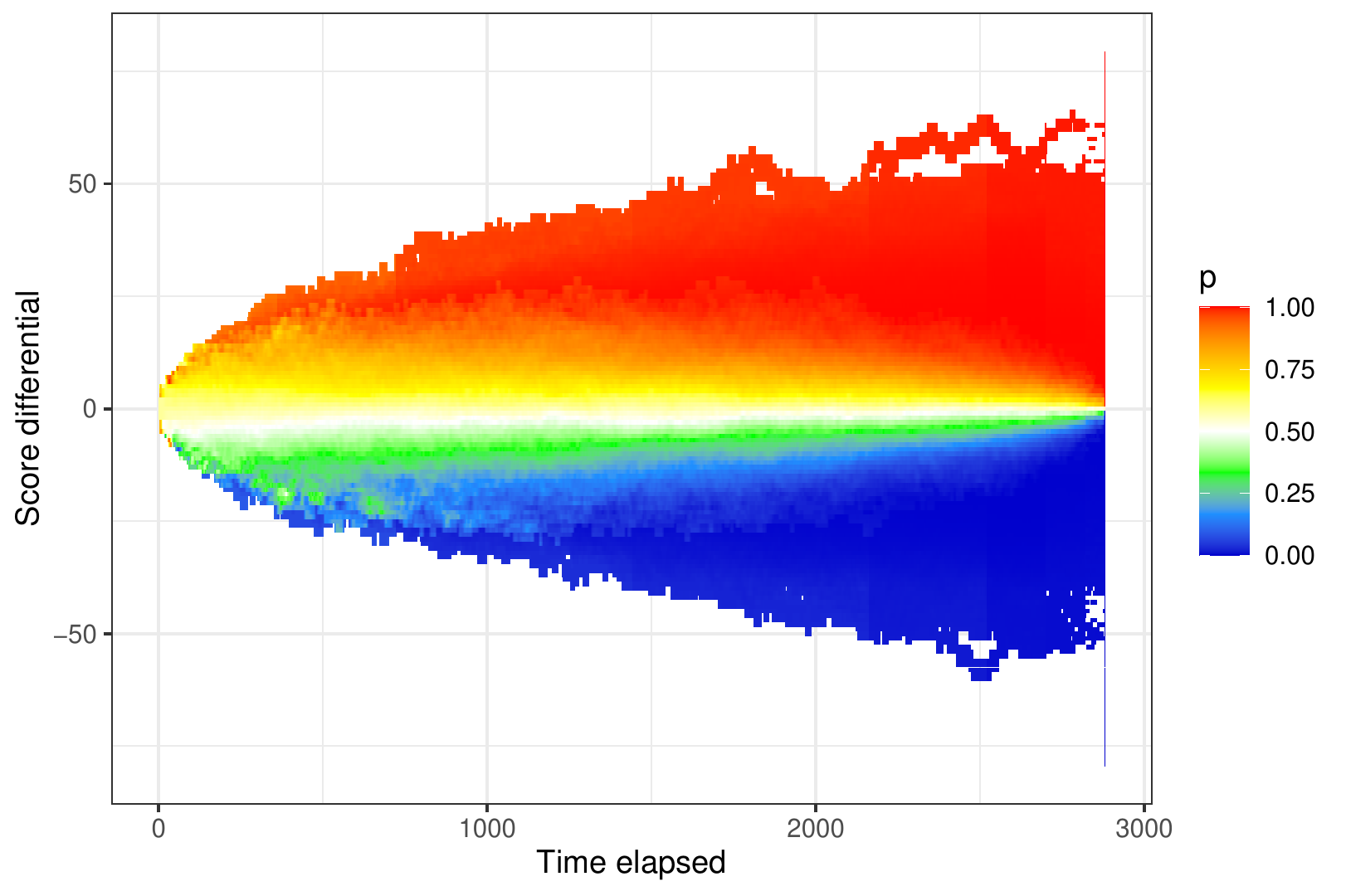} 
    \label{fig:dynbayes}
  \end{subfigure}
  \label{fig:probests}
\end{figure}

Figure~\ref{fig:probests} shows the maximum likelihood estimates (MLE) $\bar p_{t, \ell}$ from equation~\eqref{eq:MLE} and the Bayesian estimates $\hat p_{t,\ell}$ from equation~\eqref{eq:bayes} for the same data. The graph of the MLE is not as smooth as the graph of the Bayesian estimates.  This is easily explained by noting that the MLE are computed on each $(t, \ell)$ point, and not across a rectangular bin.  There are more points $(t, \ell)$ with no games (and so no estimate) than rectangular bins with no games.  Moreover, since each MLE is computed based on the number of games at point $(t, \ell)$, the standard error of the MLE is likely to be greater than the corresponding Bayesian estimate.   

The Bayesian prior parameters have an interesting interpretation, first noted by~\citet{deshpande_jensen_2016}.  The parameter $\alpha_{t, \ell}$ can be interpreted as the number of ``pseudo-wins'' in that cell; likewise $\beta_{t, \ell}$ as the number of ``pseudo-losses.''  Through this interpretation, the two parameteres can be seen as a way of increasing the number of games in a specific $(t, \ell)$ cell.  If the home team is ahead, then first scale parameter being large effectively acts to increase the number of wins in that cell.  On the other hand, if the home team is behind, the second scale parameter is large, and acts to increase the number of losses.  These ideas contribute to the smoother appearance of the Bayesian estimates.

\begin{table}[!ht]
    \caption{Imputed parameters for beta prior}
    \label{tab:dynamicprior} 
    \begin{tabular}{||l|l|cc||} \hline\hline
      \multicolumn{1}{||c|}{Elapsed Time ($t$)} & \multicolumn{1}{c|}{Home Team} & & \\ 
      \multicolumn{1}{||c|}{(sec.)} & \multicolumn{1}{c|}{Lead ($\ell$)} & $\alpha_{t, \ell}$ & $\beta_{t, \ell}$ \\ \hline
      0 -- 360 & 0 -- 9 & 1 & 1 \\
      0 -- 360 & 10 -- 14 & 18 & 9 \\
      0 -- 360 & $\geq$ 15 & 54 & 6 \\
      361 -- 720 & 0 -- 9 & 1 & 1 \\
      361 -- 720 & 10 -- 19 & 19 & 7 \\
      361 -- 720 & $\geq$ 20 & 34 & 3 \\
      721 -- 1440 & 0 -- 9 & 1 & 1 \\
      721 -- 1440 & 10 -- 19 & 18 & 5   \\
      721 -- 1440 & $\geq$ 20 & 51 & 2 \\
      1441 -- 2160 & 0 -- 9 & 1 & 1 \\
      1441 -- 2160 & 10 -- 14 & 22 & 6 \\
      1441 -- 2160 & 15 -- 19 & 15 & 2  \\
      1441 -- 2160 & $\geq$ 20 & 71 & 2 \\
      2161 -- 2520 & 0 -- 9 & 1 & 1 \\
      2161 -- 2520 & 10 -- 14 & 22 & 3 \\
      2161 -- 2520 & 15 -- 19 & 25 & 2 \\
      2161 -- 2520 & $\geq$ 20 & 133 & 2 \\
      2521 -- 2700 & 0 -- 9 & 1 & 1 \\
      2521 -- 2700 & 10 -- 14 & 46 & 3 \\
      2521 -- 2700 & 15 -- 19 & 48 & 1 \\
      2521 -- 2700 & $\geq 20$ & 133 & 1 \\
      2701 -- 2820 & 0 -- 4 & 1 & 1 \\
      2701 -- 2820 & 5 -- 9 & 10 & 2 \\
      2701 -- 2820 & 10 -- 14 & 104 & 3 \\
      2701 -- 2820 & $\geq$ 15 & 328 & 2 \\
      2821 -- 2879 & 0 -- 2 & 1 & 1 \\
      2821 -- 2879 & 3, 4 & 10 & 2 \\
      2821 -- 2879 & 5 -- 9 & 17 & 1 \\
      2821 -- 2879 & $\geq$ 10 & 167 & 1 \\ \hline\hline
    \end{tabular}
\end{table}
\clearpage 

\subsection{Binning Procedure for NBA Games}
\label{sec:bins}
The approach in~\citet{maddoxetal_2022} fixes bin area; in other words, regardless of time remaining and score differential, the length and width of all bins are equal.  However there are practical issues with fixed area binning, especially closer to the end of the game.  Consider that during the middle of the game, a two point difference in score should not have a major effect on the win probability.  However, closer to the end of a close game a two point difference in score could have a very large effect on win probability. For example, suppose there are five seconds remaining and a tie score with a jump ball to take place. Intuitively there should be about a 50\% probability that either team wins the game. However, with the same amount of time if one of those teams is up by two points, that team should have a greater than 50\% probability of winning.  In this scenario, it is not reasonable to use the same bin width on score differential at the end of the game as at the beginning.  We propose that between 2700 and 2820 seconds into the game, or three to one minutes remaining, the width of the score differential bins is shortened to $[\ell - 1, \ell + 1]$, and after 2820 seconds into the game, or in the last minute, there is no binning on score differential.  The bin widths on time remain the same at the end of the game, because with a specific score differential a small shift in time should not have a large effect on win probability even at the end of the game.

\section{Adjusted Dynamic Bayesian Estimator}
\label{sec:adjbayes}
In-game win probability is certainly a function of time and score differential during the game.  However, it is also affected by the skill of the teams playing the game.  Incorporating some measure of team ability in to the model was also discussed in~\citet{maddoxetal_2022}.  The normal distribution quantile function was used to convert the pregame point to a pregame home team win probability $\hat p_p$ for each team.  The pregame probability, $\hat p_p$, was added into the model for predicting in-game win probability so that the weight of $\hat p_p$ decreased linearly as a function of time remaining in the game to get a final adjusted Bayesian estimate of in-game home-team win probability.  In what follows, the model for finding $\hat p_p$ and including it in the model is refined by allowing for more complex models for including $\hat p_p$.  The resulting final ``adjusted dynamic Bayesian'' estimates are compared with the non-adjusted dynamic Bayesian estimates and the ESPN counterparts. 

\subsection{Brier's Score}
\label{sec:briersscore}
Brier's score is a statistic used to compare the performance of different methods for estimating probabilities.  Brier's score is the average of the square of the difference between the estimated probability and the observed binary outcome.  In the context of in-game home team win probabilities, this observed binary outcome is denoted $y_i$ and is the observed value of $Y_i$ as defined in Section~\ref{sec:model}.  To interpret Brier's score, if $Y_i = 1$ for all $i$, and the predicted probability is also one for every $i$, then Brier's score will be zero, indicating perfect prediction.  On the other hand, if for all $i, Y_i = 0$, and the estimated probability is one, then Brier's score will be one, the worst possible Brier's score.  

To compute Brier's score, if $\rho_{t, \ell}$ represent the estimated in-game home team win probability.  For each $(t, \ell)$ cell, let $N^*_{t,\ell}$ represent the number of games in the cell in which the home team led by $\ell$ points at time $t$; that is, $N^*_{t,\ell}$ is the number of games observed in that cell.  Then for non-missing estimated probabilities, Brier's score is
\begin{equation*}
B = \frac{1}{Q}\sum_{t=0}^{2879}\sum_{\ell=-58}^{58} \sum_{j=1}^{N^*_{t,l}} \left(\rho_{t,\ell} - y_j\right)^2,
\end{equation*}
where $Q$ is the sum of $N^*_{t,\ell}$ in cells without missing $\rho_{t,\ell}$.
When evaluating the models using Brier's Score, the 2018-19 and 2019-20 seasons are used as testing data, each having $Q = 6,590,576$ observations.

\subsection{Pregame Win Probabilities}
\label{sec:pregamewinprobs}
\citet{maddoxetal_2022} adjustment to dynamic Bayesian estimator by including a measure of pregame win probabilities, linearly shifting the weight from pregame win probabilities to in-game win probabilities across using
\begin{equation}
    \label{eq:best}
    \hat p^*_{t, \ell} = \left(\frac{S - t}{S}\right)\hat p_p + \left(\frac{t}{S}\right)\hat p_{t, \ell},
\end{equation}
where $S$ is the number of seconds in the game, and is 2,880 for NBA games\footnote{In~\citeauthor{maddoxetal_2022}, $S = 2,400$, the number of seconds in the NCAA basketball game.}, $\hat p_{t, \ell}$ is given in equation~\eqref{eq:bayes} using the binning procedure described in Section~\ref{sec:bins} and the prior structure given in Table~\ref{tab:dynamicprior}, $\hat p^*_{t, \ell}$ is the final predicted win probability and $\hat p_p$ is the pregame win probability.  Incorporating pregame win probabilities into the model helps to improve predictive accuracy because team quality plays a role in predicting who may win a game.  The linear adjustment is simple, and only a function of time remaining. 

In what follows, we investigate three different functions for incorporating pregame probabilities.  The first is a linear function of only time remaining.  The second is a linear function of time remaining and score differential.  The third is linear in time, but quadratic in score differential.  Other variations to the three weight functions were considered. For example, including a quadratic term for time was attempted. However, the more complicated models would not converge appropriately for \emph{R} to optimize the performance accurately.  

Let the $A$ be some set of numbers.  Then for a specific number $a$, the indicator function ${\mathbbm{1}}_A(a)$ is defined as
$$
{\mathbbm{1}}_A(a) = \begin{cases}
1 & {\mbox{if $a \in A$, and}} \\
0 & {\mbox{otherwise.}}
\end{cases}
$$
The three weight functions specifically considered here are
\begin{align*}
B_1 & = b t, \\
B_2 & = c_1 t + c_2|\ell|, \\
B_3 & = d_0 + d_1{\mathbbm{1}}_0(\ell) + d_2 t + d_3 |\ell| + d_4\ell^2.
\end{align*}
Each one of these weight functions yields a competing model for including pregame win probabilities, which can be represented
\begin{equation*}
    p^*_{t, \ell, j} = \begin{cases} \hat p_p, & B_j \leq 0 \\ (1 - B_j) \hat p_p + B_j \hat p_{t, \ell}, & 0 < B_j < 1 \\ \hat p_{t, \ell} & B_j \geq 1 \end{cases}, \qquad j = 1, 2, 3.
\end{equation*}
The model for $p^*_{t, \ell, 1}$ is equivalent to the model in equation~\eqref{eq:best} for $b = 1/2880$ of~\citeauthor{maddoxetal_2022}.  The weight function $B_2$ includes a linear dependence on both time and score differential.  Additionally, if the value of $B_2$ is greater than one, only $p^*_{t, \ell, 2} = \hat p_{t, \ell}$, the Bayesian estimator in equation~\eqref{eq:bayes}.  Finally, $p^*_{t, \ell, 3}$ specifies a linear effect in time remaining, but quadratic effect of score differential as well as an intercept term, $d_0$.  As with $B_2$, if $B_3 > 1, p^*_{t, \ell, 3} = \hat p_{t, \ell}$.  Since $\ell$ enters $B_3$ through only $|\ell|$ and $\ell^2, B_3$ is more likely to be greater than 1.  

Values for $b, c_1, c_2, d_0, d_1, \ldots, d_4$ are estimated by choosing those values which minimize Brier score for each $p^*_{t, \ell, j}$ tested on data from the 2018--19 and 2019--20 seasons. The Brier score for each $p^*_{t, \ell, j}$ is shown in Table~\ref{tab:pro_brier}. The model $p^*_{t, \ell, 3}$ is the most accurate predictor when using seasons 2018-20 to evaluate the model built on seasons 2012-2018.  The fitted model for $B_3$ that provided the minimal Brier score is
$$
B_3 = -1.10633 - 0.02313{\mathbbm{1}}_{0}(\ell) + 0.00027t + 0.06618|\ell| - 0.00139\ell^2.
$$
Using this expression for $B_3$, the model
$$
    p^*_{t, \ell, 3} = \begin{cases} \hat p_p & B_3 \leq 0\\ (1 - B_3) \hat p_p + B_3 \hat p_{t, \ell}, & 0 < B_3 < 1\\ \hat p_{t, \ell}, & B_3 \geq 1 \end{cases}
$$
is the ``adjusted dynamic Bayesian estimator.''

\begin{table}[!ht]
\caption{Brier Scores for models determining pregame probability proportion.}
\label{tab:pro_brier}
\begin{tabular}{||r|c||} \hline\hline
Proportion Model                & Brier Score \\ \hline
Linear Time           $(B_1)$   & 0.1622       \\
Linear Time \& Score  $(B_2)$   & 0.1613       \\
Quadratic             $(B_3)$   & 0.1598       \\ \hline\hline
\end{tabular}
\end{table}

\begin{figure}[!ht]
\caption{Adjusted dynamic Bayesian estimates for different values of pregame win probability.}
  \begin{subfigure}[h]{0.325\textwidth}
    \caption{$\hat p_p = 0.59$.}  
    \includegraphics[width=\textwidth]{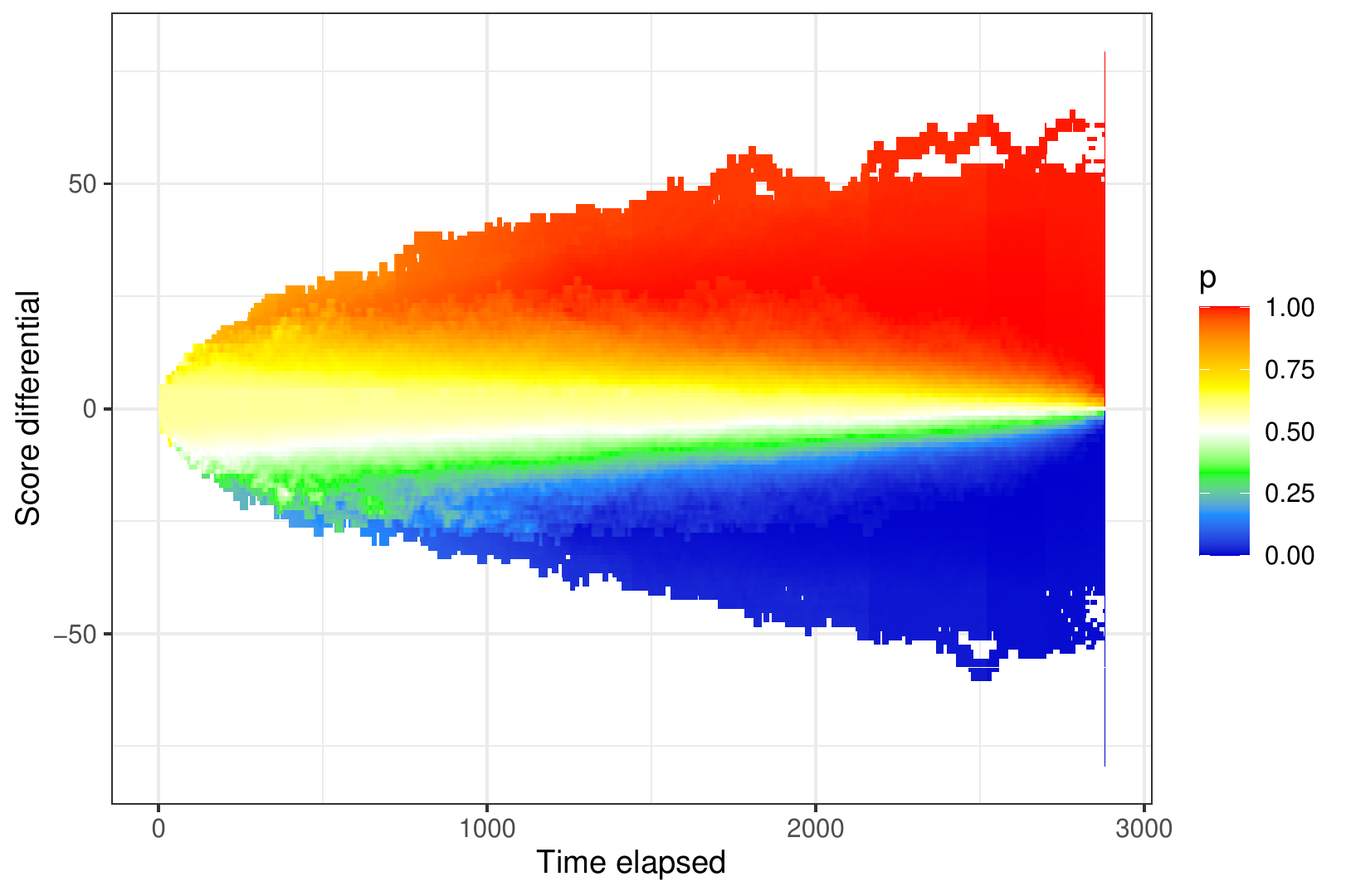}
    \label{fig:aveadjBayes}
  \end{subfigure}
  \hfill
  \begin{subfigure}[h]{0.325\textwidth}
    \caption{$\hat p_p = 0.89$.}
    \includegraphics[width=\textwidth]{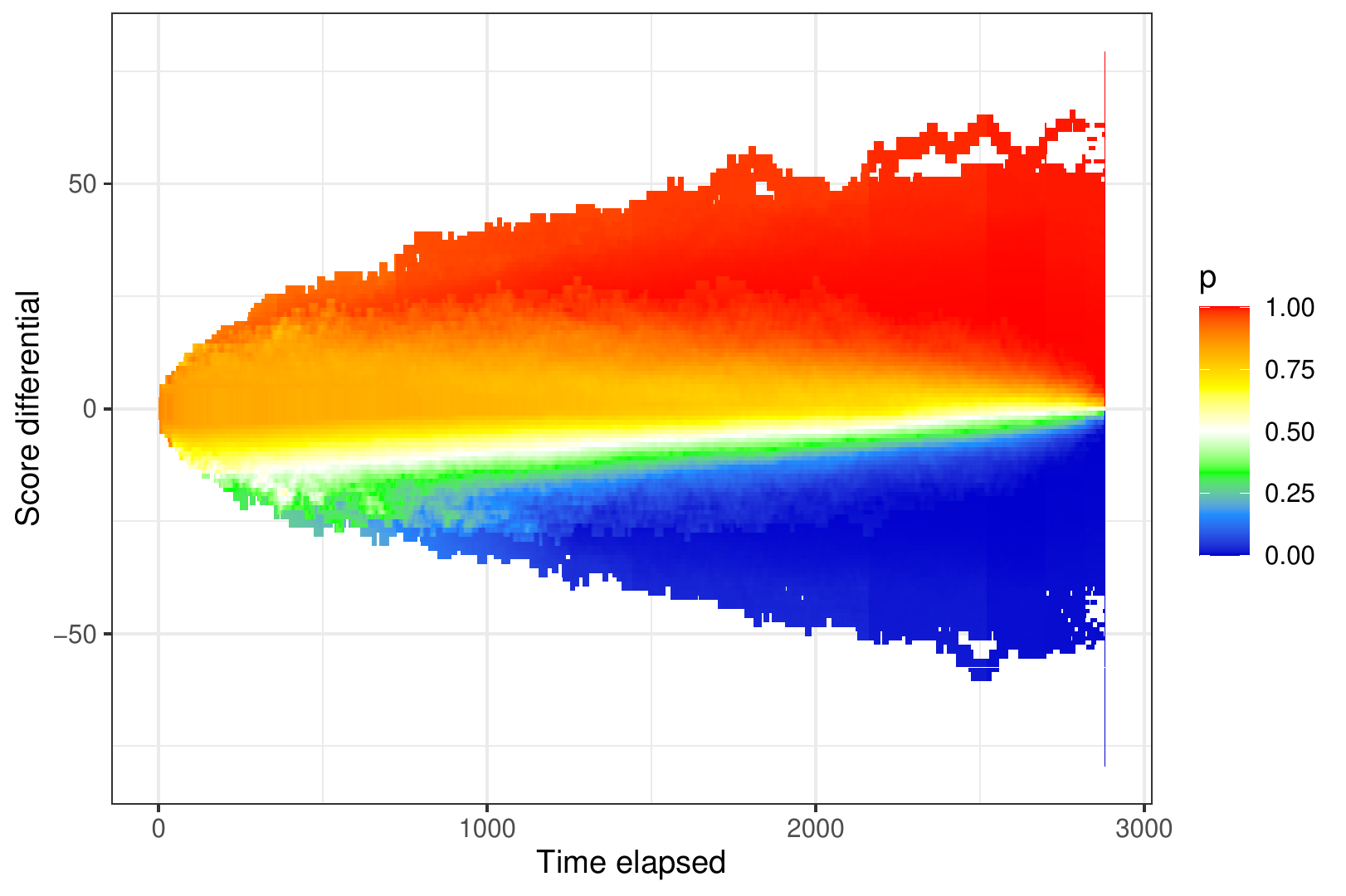}
    \label{fig:homeadjBayes}
  \end{subfigure}
  \hfill
  \begin{subfigure}[h]{0.325\textwidth}
    \caption{$\hat p_p = 0.29$.}
    \includegraphics[width=\textwidth]{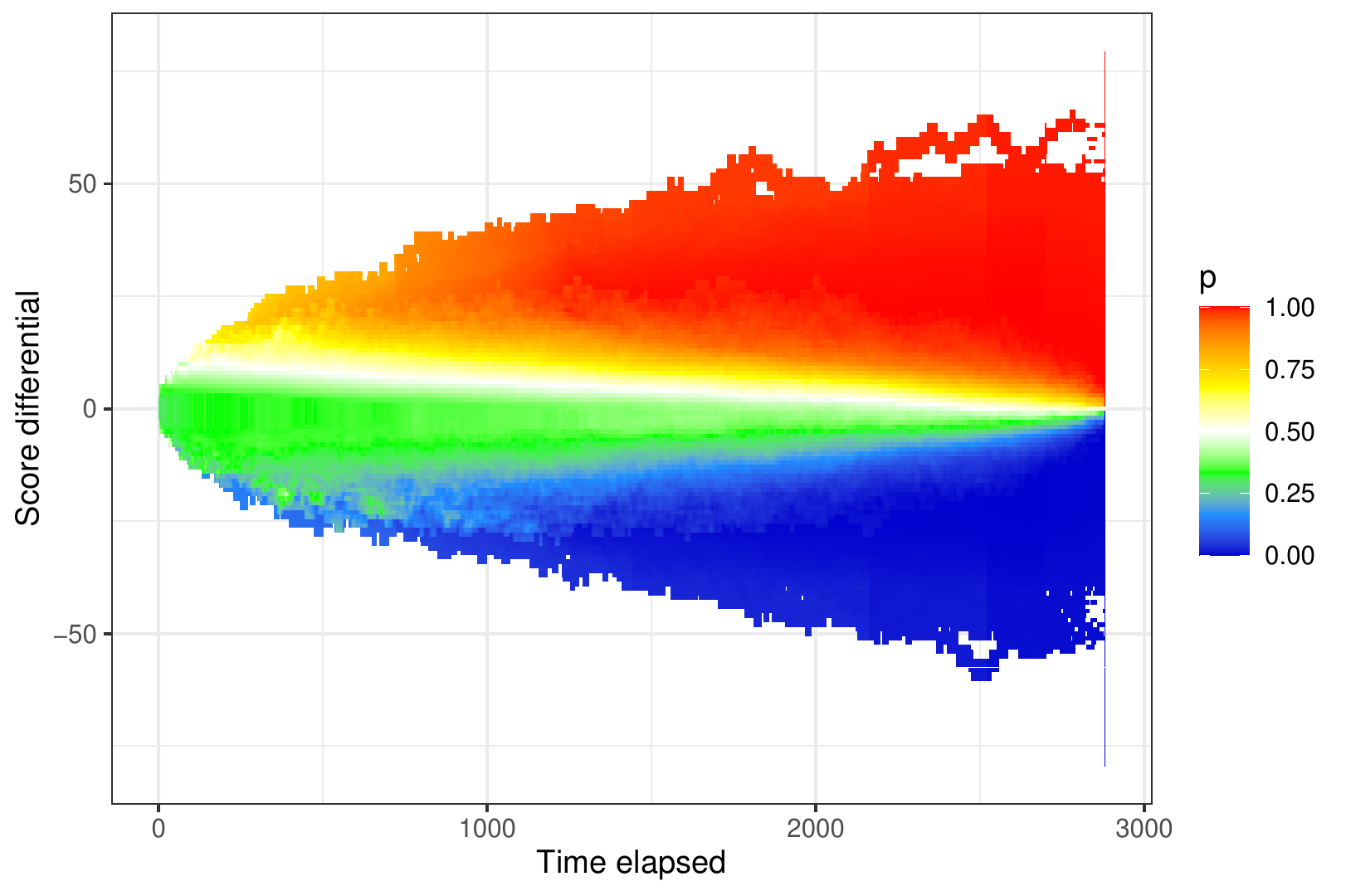}
    \label{fig:roadadjBayes}
  \end{subfigure}
\end{figure}

Figures~\ref{fig:aveadjBayes} through~\ref{fig:roadadjBayes} illustrate the estimated in-game home team win probabilities using the adjusted Bayes estimator, $p^*_{t, \ell, 3}$.  Because the estimated probability is affected by the pregame home team win probability, three values of $\hat p_p$ were selected to illustrate the performance of the newly proposed estimator.  Of the 7,376 games in the training data, approximately 59\% were won by the home team, motivating the choice of $\hat p_p = 0.59$.  The other two choices of pregame home team win probability are $\hat p_p = 0.59 \pm 0.3$. 

\subsection{Model Comparison}
\label{sec:performance}
For each second of each game from the 2018-19 and 2019-20 seasons, 
Brier's score was computed.  The Brier Scores for each of the three models are shown in Table~\ref{tab:mod_brier}.  As mentioned previously, ESPN's win probability model is not publicly available, and the model itself cannot be reproduced. However, ESPN does release the results of the model, which were scraped from their website.  Using Brier Scores, ESPN's win probability model does outperform the new adjusted dynamic Bayesian model for both seasons in the test data set. However, Table~\ref{tab:ingame-eval-brier} shows the performance of each model at different times during the game. The adjusted dynamic Bayesian model performs better than the ESPN model at several points in the fourth quarter in both seasons. This may indicate that future improvements could be made to the adjusted dynamic Bayesian model by adjusting the pregame portion of the model.

\begin{table}[!ht]
    \caption{Brier Scores for predictive performances for the 2018-2019 and 2019-2020 seasons.}
\begin{tabular}{||l|c|c|c||} \hline\hline 
                                        & 2018-2019 Season  & 2019-2020 Season  &   Total  \\ \hline
    Bayes with dynamic prior            & 0.1663            & 0.1736            &   0.1697  \\
    Adjusted dynamic Bayes              & 0.1568            & 0.1635            &   0.1598   \\
    ESPN win probabilities              & 0.1550            & 0.1621            &   0.1582    \\ \hline\hline
    \end{tabular}
    \label{tab:mod_brier}
\end{table}

\begin{table}[!ht]
\caption{In-game evaluation of predictive performances for the 2018-2019 and 2019-2020 seasons.}
\label{tab:ingame-eval-brier}
\scriptsize
\begin{tabular}{||r|ccccc|ccccc||} \hline\hline
 & \multicolumn{5}{c|}{2018-2019 Season} & \multicolumn{5}{c||}{2019-2020 Season} \\ \cline{2-11}
 & \multicolumn{10}{c||}{Time remaining (in minutes)} \\ \cline{2-11}
 & 24 & 12 & 6 & 3 & 1 & 24 & 12 & 6 & 3 & 1 \\ \cline{2-11}
Dyn Bayes                      & 0.1761 & 0.1194 & 0.0935 & 0.0775 & 0.0537 & 0.1840 & 0.1334 & 0.1071 & 0.0822 & 0.0588 \\
Adj Dyn Bayes                   & 0.1702 & 0.1168 & 0.0927 & 0.0773 & 0.0537 & 0.1762 & 0.1300 & 0.1063 & 0.0819 & 0.0588 \\
ESPN                    & 0.1666 & 0.1181 & 0.0931 & 0.0769 & 0.0536 & 0.1738 & 0.1311 & 0.1079 & 0.0829 & 0.0584 \\ \hline\hline
\end{tabular}
\end{table}

\subsection{{\tt TeamRankings.com} vs.~Elo}
\label{sec:phatpchoices}
While various metrics can be used to arrive at pregame win probabilities, short of adjusting for starting lineups, power rankings should work well in determining the pregame win probabilities. Metrics from \url{TeamRankings.com} might not be superior to others.  However, many others do not have public daily ratings.  Another common and well-researched win probability system, called the Elo ranking system, is often used to predict team performance for a game which can be used to compare the differences between various power ranking systems. The website \url{FiveThirtyEight.com} provides historical Elo ratings for each team for each day they played a game dating back to the 1946-47 season. 

We can compare the performance of \url{TeamRankings.com} win probabilities and Elo win probabilities by both comparing their Brier scores or by comparing their performance when applied to the model introduced in this paper.  These results are found in Table~\ref{tab:elo}. The lower Brier score for \url{TeamRankings.com} pregame predictions compared to that of the Elo model shows \url{TeamRankings.com} performs better overall.  Additionally, when the Elo pregame win probabilities are applied to the methodology outlined in this paper the Brier score for the adjusted dynamic Bayesian model is 0.1598 using \url{TeamRankings.com} compared to 0.1605 for Elo.
\begin{table}[!ht]
    \caption{Brier Scores for \url{TeamRankings.com} compared to Elo.}
    \begin{tabular}{||l|c|c||} \hline\hline
        Pregame probability model           &   Pregame model   &   In-game model   \\  \hline
        TeamRankings.com                    &   0.2167          &   0.1598          \\
        Elo                                 &   0.2179          &   0.1605          \\ \hline\hline
    \end{tabular}
    \label{tab:elo}
\end{table}

\section{Application to a Game}
\label{sec:application}
\begin{figure}[!ht]
  \caption{In-game win probabilities for Chicago at Charlotte.}
  \includegraphics[width=\textwidth]{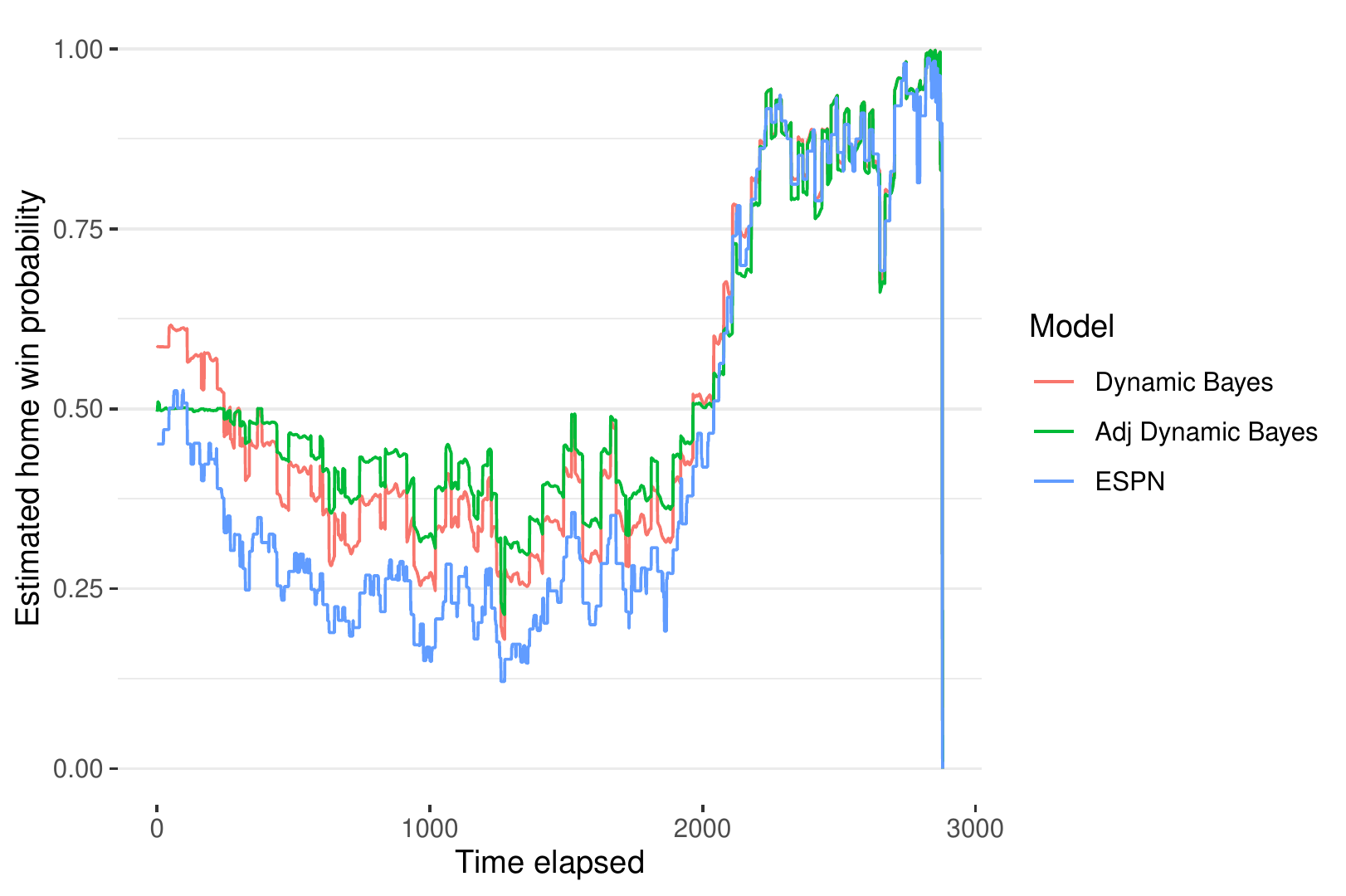}
  \label{fig:graph}
\end{figure}

On  November 23, 2019 the Chicago Bulls travelled to play at the Charlotte Hornets, in what was expected to be a trivial, low-profile regular season game between two relatively unsuccessful teams. However, the game quickly turned into one of the most exciting games of the season, with Zach LaVine setting a career high of 49 points, including a game-winning 3-point shot with less than one second left in the game\footnote{Since that game, LaVine scored a new career high 50 points on April 9, 2021 against the Atlanta Hawks.}. The Bulls ended the game on a 16-7 run in the last minute of the game. The win probabilities for this game produced from each of the three models are displayed in Figure~\ref{fig:graph}. The largest differences in estimated win probability for the three models occur early on during the game. This is to be expected, as early on in the game there will be more variance in win probability due to more time allowing for many possible unpredictable occurrences in the game. Throughout the first three quarters of the game, ESPN predicts a larger probability of the Bulls winning the game than the other two models. However, many of the rises and falls in the three models occur at the same time in the game. During the fourth quarter, the three models are close to indistinguishable from each other, each giving very similar results for this particular game. Lastly, at the end of the game all three models had an extremely high probability that the Hornets would win, signifying the remarkable run the Bulls went on in the last minute, culminating in a buzzer beating three point shot that swung all three models from predicting a Hornets' win to a Bulls' win.  The graph shows this by what appears to be an almost perfectly vertical blue line at the end of the game.  However, that line is actually all three models having home win probability close to 1, then plummeting to zero once LaVine's shot is made.

\section{Conclusion}
\label{sec:conclusion}
Two new methods are proposed for estimating in-game win probability for NBA games. Both are an extension and enhancement of the methods in~\citet{maddoxetal_2022}, which provide a number of models for estimating in-game home-team win probabilities for NCAA basketball.  The first proposed ``dynamic Bayesian estimator,'' uses a prior that has been calculated based on the distribution of predicted win probabilities from 14 NBA field experts, including several anonymous front office associates within the NBA. The second method, referred to as the ``adjusted dynamic Bayesian estimator,'' adjusts the dynamic Bayesian estimator based on pregame win probabilities obtained from \url{TeamRankings.com}. The adjustment is optimized over a function of both time and score so that as the game moves on or the score differential increases, the adjusted dynamic Bayesian estimator will begin to approach the dynamic Bayesian estimator rather than the pregame win probability. These two methods are then compared to the win probability model that ESPN uses for seasons 2018-19 and 2019-20. The ESPN model performs the best overall, but there are times during the game the adjusted dynamic Bayesian model performs the best, indicating that there are some features of the model that estimate probabilities well and some that could be improved upon in the future, such as the calculation of pregame win probability.

\bibliographystyle{apalike}
\bibliography{refs}
\end{document}